\documentclass[aps,prl,groupedaddress,%
  twocolumn,%
  showpacs,floatfix
]{revtex4}

\usepackage{amsmath}
\usepackage{amssymb}
\usepackage{txfonts}
\usepackage{graphicx}

\DeclareMathOperator{\dn}{dn}
\DeclareMathOperator{\sech}{sech}
\renewcommand{\L}{\mathbf{L}} \newcommand{\N}{\mathbf{N}}
 \newcommand{\F}{\mathcal{F}}
\renewcommand{\epsilon}{\varepsilon}
\newcommand{\caseI}{%
  \vrule height6pt width7pt depth-5.6pt %
  \vrule height6pt width0.4pt depth-2.1pt %
  \vrule height2.5pt width7pt depth-2.1pt %
  \vrule height2.5pt width0.4pt depth1pt %
  \vrule height-0.6pt width7pt depth1pt %
  }
\newcommand{\caseII}{%
  \vrule height6pt width7pt depth-5.6pt %
  \vrule height6pt width0.4pt depth1pt %
  \vrule height-0.6pt width7pt depth1pt %
  \vrule height2.5pt width0.4pt depth1pt %
  \vrule height2.5pt width7pt depth-2.1pt %
  }
\newcommand{\caseIII}{%
  \vrule height2.5pt width7pt depth-2.1pt %
  \vrule height6pt width0.4pt depth-2.1pt %
  \vrule height6pt width7pt depth-5.6pt %
  \vrule height6pt width0.4pt depth1pt %
  \vrule height-0.6pt width7pt depth1pt %
  }
\newcommand{\caseIV}{%
  \vrule height2.5pt width7pt depth-2.1pt %
  \vrule height2.5pt width0.4pt depth1pt %
  \vrule height-0.6pt width7pt depth1pt %
  \vrule height6pt width0.4pt depth1pt %
  \vrule height6pt width7pt depth-5.6pt %
  }
\newcommand{\caseV}{%
  \vrule height-0.6pt width7pt depth1pt %
  \vrule height6pt width0.4pt depth1pt %
  \vrule height6pt width7pt depth-5.6pt %
  \vrule height6pt width0.4pt depth-2.1pt %
  \vrule height2.5pt width7pt depth-2.1pt %
  }
\newcommand{\caseVI}{%
  \vrule height-0.6pt width7pt depth1pt %
  \vrule height2.5pt width0.4pt depth1pt %
  \vrule height2.5pt width7pt depth-2.1pt %
  \vrule height6pt width0.4pt depth-2.1pt %
  \vrule height6pt width7pt depth-5.6pt %
  }

\begin{document}

\title{Soliton Generation and Multiple Phases in 
  Dispersive Shock and Rarefaction Wave Interaction}  

\author{M.\ J.\ Ablowitz}
\author{D.\ E.\ Baldwin}
\email[]{shockwaves@douglasbaldwin.com}
\affiliation{Department of Applied Mathematics, 
  University of Colorado, Boulder, Colorado, 80309-0526, USA}
\author{M.\ A.\ Hoefer}
\affiliation{Department of Applied Physics and Applied Mathematics, 
  Columbia University, New York, New York, 10027, USA}

\date{\today}

\begin{abstract} 
Interactions of dispersive shock (DSWs) and rarefaction waves (RWs) associated 
with the Korteweg-de Vries equation are shown to exhibit multiphase dynamics 
and isolated solitons.  There are six canonical cases: one is the 
interaction of two DSWs which exhibit a transient two-phase solution, but 
evolve to a single phase DSW for large time; two tend to a DSW with either a 
small amplitude wave train or a finite number of solitons, which can be 
determined analytically; two tend to a RW with either a small wave train or a 
finite number of solitons; finally, one tends to a pure RW.
\end{abstract}

\pacs{47.40.Nm,05.45.Yv,52.35.Mw}

\maketitle

Shock waves in processes dominated by weak dispersion 
and nonlinearity have been experimentally observed in 
plasmas~\cite{Taylor1970}, water waves~\cite{Smyth1988}, 
and more recently in Bose-Einstein 
condensates~\cite{Hoefer2006,Chang2008} 
and nonlinear optics~\cite{Wan2007}; 
these dispersive shock waves (DSWs) have yielded novel dynamics 
and interesting interaction behavior which has only recently begun 
to be studied theoretically (cf.\ \cite{El2002,Hoefer2007}).  
Here we consider DSWs 
which are described by the Korteweg-de Vries (KdV) equation, 
\begin{equation} \label{eq:KdV}
  u_t + uu_x + \epsilon^2 u_{xxx} = 0, 
  \qquad 0 < \epsilon \ll 1. 
\end{equation}
Individual DSWs are characterized by a soliton train front 
with an expanding oscillatory wave at its trailing edge; 
these waves have been well-studied 
(cf.\ \cite{Gurevich1974,avgTheory}) 
using wave averaging techniques, often referred to as Whitham 
theory~\cite{Whitham1965,Whitham1974}.  

When illustrative, 
we contrast DSW interaction with classical or viscous shock waves (VSWs), 
which are dominated by weak \emph{dissipation} and nonlinearity,
using Burgers' equation
\begin{equation} \label{eq:Burgers}
  u_t + uu_x - \nu u_{xx} = 0, \qquad 0 < \nu \ll 1.
\end{equation}  
The interaction of VSWs is an entire field and has been extensively studied 
(cf.\ \cite{vswTheory}), while little is known about DSW interactions.  

In this letter, we use analytic, asymptotic and numeric 
methods to investigate (\ref{eq:KdV}) (and (\ref{eq:Burgers})) 
using the ``step-like'' initial data 
\begin{equation} \label{eq:uIC}
  u(x,0) = u_0(x) = 
  \begin{cases} 
    h_0, & x < 0, \\
    h_1, & 0 < x < L, \\
    h_2, & x > L,
  \end{cases}
\end{equation}
where $h_0$, $h_1$ and $h_2$ are distinct, real 
and non-negative.  
This gives six canonical cases, which we denote:
\[\begin{aligned}
  \text{I (\caseI):} \hspace{1ex} & h_0 > h_1 > h_2, 
  \quad& \text{II (\caseII):} \hspace{1ex} & h_0 > h_2 > h_1, \\ 
  \text{III (\caseIII):} \hspace{1ex} & h_1 > h_0 > h_2,
  \quad& \text{IV (\caseIV):} \hspace{1ex} & h_2 > h_0 > h_1, \\
  \text{V (\caseV):} \hspace{1ex} & h_1 > h_2 > h_0,
  \quad&  \text{VI (\caseVI):} \hspace{1ex}& h_2 > h_1 > h_0,
  \end{aligned} \]
where an icon 
of the initial step data is shown in parentheses.  
When convenient, and without loss of generality, 
we take $h_i$ to be $0$, $1$ and $0 < h_* < 1$ 
(by using a scaling symmetry and Galilean invariance). 
The case of an initial depression 
(e.g. Case II, $h_0 = h_2 = 0 > h_1$) 
and an initial box (e.g., Case III, $h_0 = h_2 = 0 < h_1$) 
has been studied in \cite{El2002}, 
where the asymptotic solution was constructed 
analytically.  

This letter is organized as follows.  
We first discuss Case~I (\caseI), 
where two DSWs interact and exhibit a two-phase region  
which evolves into effectively a one-phase solution for large time.  
Single phase Whitham theory is then introduced to 
describe the DSW with a small amplitude wave train 
which develops in Case~II (\caseII).  
We then briefly discuss multiphase Whitham theory to describe 
the two-phase region in Case~I (\caseI).  
In Case~III (\caseIII), 
the interaction produces a DSW with a finite number of solitons, 
which remarkably can be determined analytically 
using Inverse Scattering Transform (IST) theory (cf.\ \cite{Ablowitz1991}).  
There is no analogue for emerging solitons in VSWs.  
We then use Whitham and IST theory to describe the 
interactions in Case~IV (\caseIV), V (\caseV) and~VI (\caseVI).
Finally, we comment on the numerical scheme we used to solve 
(\ref{eq:KdV}) and~(\ref{eq:Burgers}).  

\medskip

\begin{figure}
\halign{\hfil # \hfil & \hfil # \hfil\cr
  (a) \quad $t = 2$ & (d) \quad $t = 400$ \cr
  \includegraphics[width=1.65in]{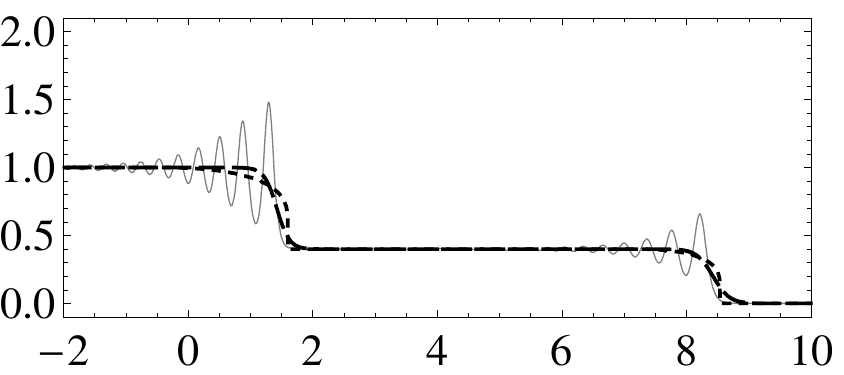} & 
  \includegraphics[width=1.65in]{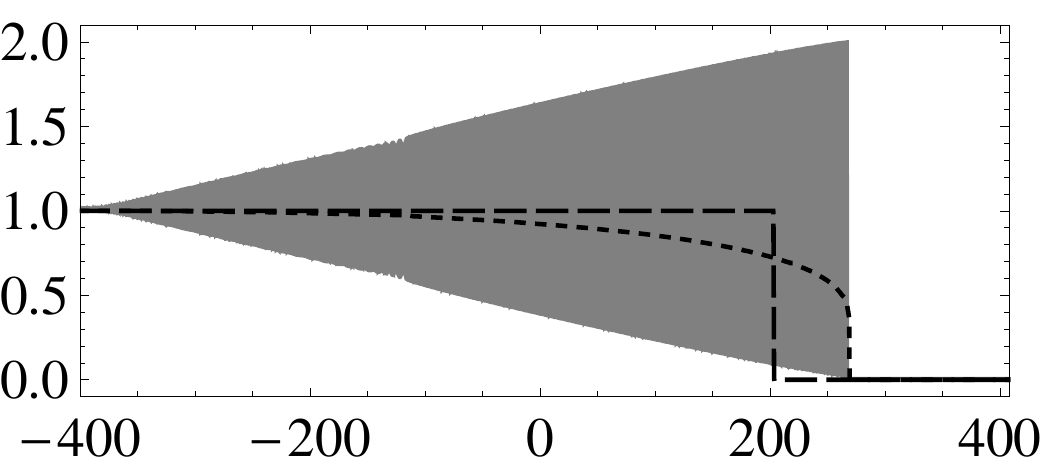} \cr
  (b) \quad $t = 10$ & (e) \cr}
\halign{\hfil # \hfil & \hfil # \hfil\cr
  \begin{minipage}[b]{1.65in}
    \halign{\hfil # \hfil \cr
    \includegraphics[width=1.65in]{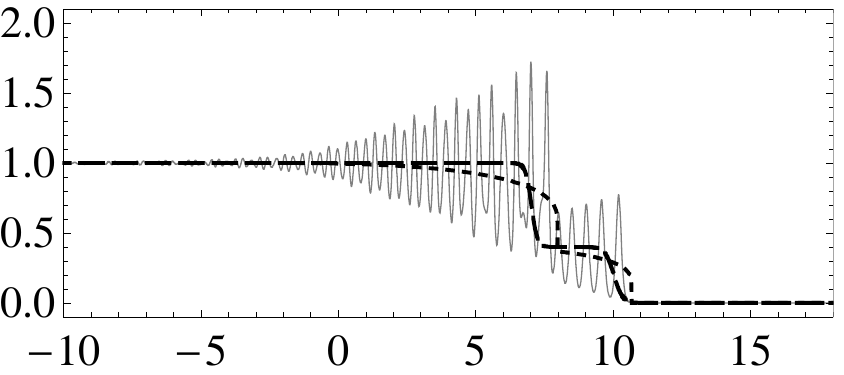} \cr 
    (c) \quad $t = 50$ \cr
    \includegraphics[width=1.65in]{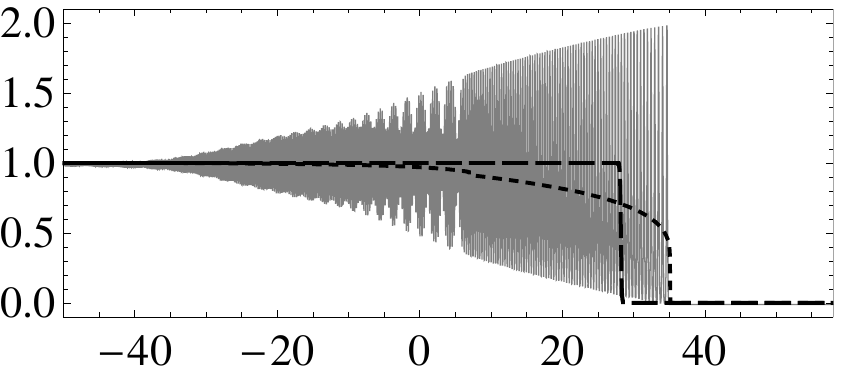} \cr}
  \end{minipage}
    \includegraphics[width=1.65in]{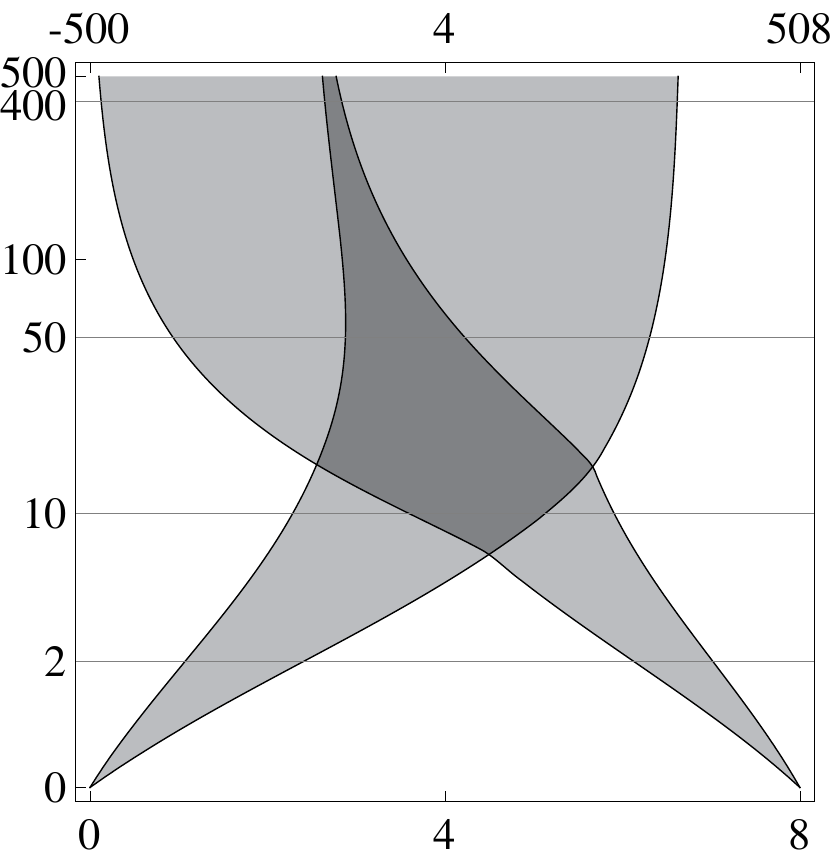} \cr}
\caption{\label{fig:C1DN} 
Plots (a)--(d) show the numerically computed solution of 
(\ref{eq:KdV}) and (e) the boundary of the one- (light gray) 
and two-phase (dark gray) regions computed using Whitham theory. 
The averaged solution, $\bar{u}$, is computed 
using Whitham averaging (cf.\ \cite{Levermore1988}) and 
shown as dotted lines in (a)--(d); 
the solution of (\ref{eq:Burgers}) is shown as dashed lines 
in (a)--(d). 
In all plots, $\epsilon^2 = 0.001$, $h_0 = 1$, $h_1 = 0.4$, 
$h_2 = 0$ and $L = 8$.  
The vertical axis in (e) is log-time and the horizontal axis is 
$-t \le x \le t+8$ (and matches the domain in (a)--(d)).  }
\end{figure}

In Case~I (\caseI), two one-phase DSWs form and propagate to the 
right (see Fig.~\ref{fig:C1DN}a).  
When the shock front of the left DSW reaches the expanding 
oscillatory tail of the right DSW, they interact and form a 
quasi-periodic two-phase solution (see Fig.~\ref{fig:C1DN}b).  
The shock front of the left DSW subsequently overtakes the 
shock front of the right DSW and forms a one-phase solution to 
the right of the two-phase region (see Fig.~\ref{fig:C1DN}c). 
To the left of the two-phase solution, 
an essentially one-phase DSW tail emerges 
(see Fig.~\ref{fig:C1DN}c); 
although the tail is weakly modulated by a quasi-periodic 
wave, its behavior is essentially one-phase.  
For large time, the two-phase region closes and 
a one-phase DSW remains (see Fig.~\ref{fig:C1DN}d--e); 
Whitham theory indicates that the amplitude of the two-phase 
modulations decrease with time and result in an effectively 
one-phase DSW. 
This closing of the two phase region is suggested by the 
rigorous (Whitham theory) results in \cite{Grava2002}, 
though the authors studied smooth initial data.  
The computation of the boundaries of the one- and two-phase 
regions using multiphase Whitham theory are discussed later in 
this letter.  

Although the (initial) shock front speed is different for DSWs 
and VSWs ($2h_0/3$ and $h_0/2$, respectively), 
the averaged DSWs are similar in behavior to VSWs 
(see Fig.~\ref{fig:C1DN}a--d); 
in both, two shock waves merge to form a single shock wave.  


\medskip

\begin{figure}
\halign{\hfil # \hfil & \hfil # \hfil\cr
  (a) \quad $t = 2$ & (c) \quad $t = 100$ \cr
  \includegraphics[width=1.65in]{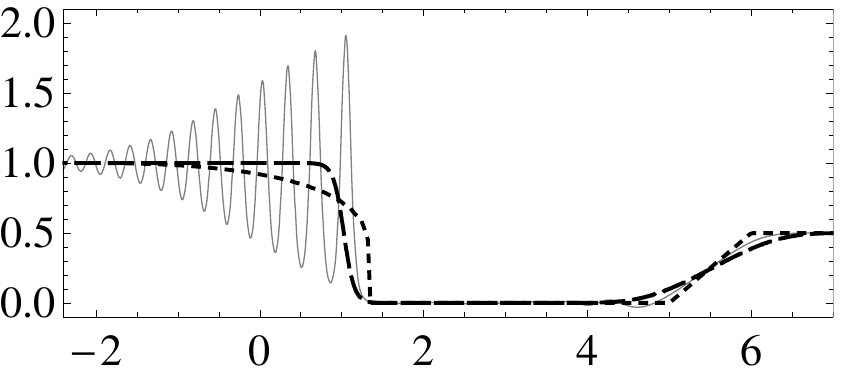} & 
  \includegraphics[width=1.65in]{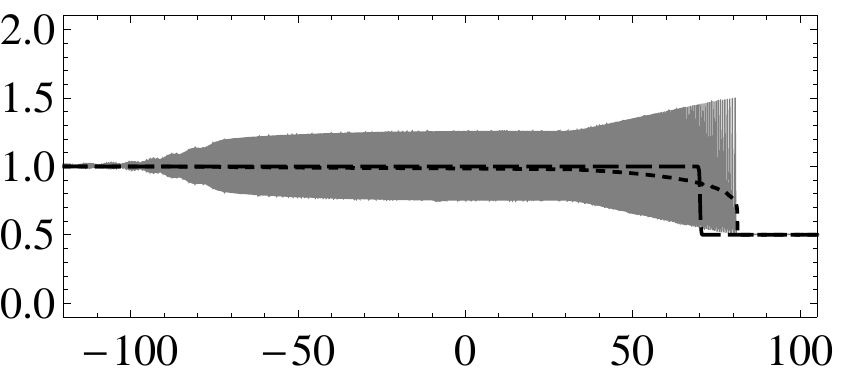} \cr
  (b) \quad $t = 10$  & (d) \quad $t = 800$ \cr
  \includegraphics[width=1.65in]{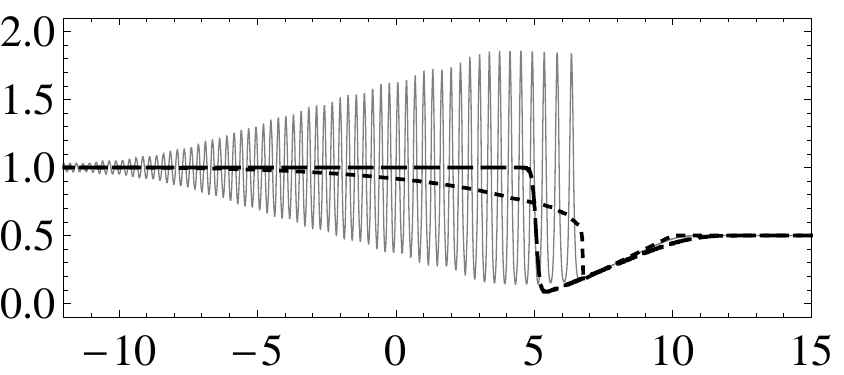} & 
  \includegraphics[width=1.65in]{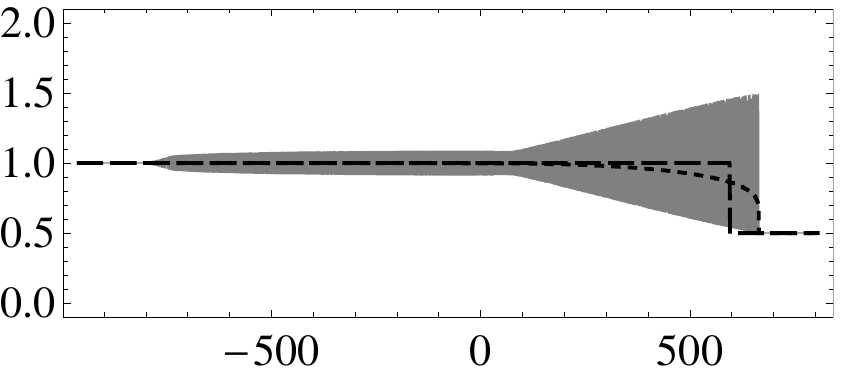} \cr}
\caption{\label{fig:C2DN} 
Plots of the numerical and averaged Whitham solutions of 
(\ref{eq:KdV}) for Case~II (\caseII) where 
$\epsilon^2 = 0.001$, $h_0 = 1$, $h_1 = 0$, $h_2 = 0.5$ and $L = 5$.}
\end{figure}

For Case~II (\caseII), 
a large DSW forms on the left and a small RW forms on the right
(see Fig.~\ref{fig:C2DN}a).  
The front of the DSW then interacts with the trailing 
edge of the RW; 
the interaction decreases the DSW's speed and height 
(see Fig.~\ref{fig:C2DN}b).
The front of the DSW is faster than the front of the RW 
and overtakes it (see Fig.~\ref{fig:C2DN}c). 
The size of the interaction region continues to expand 
with a DSW emerging in front with a small amplitude 
wave train behind, whose amplitude is proportional to 
$t^{-1/2}$ (see Fig.~\ref{fig:C2DN}d).  
As in Case~I (\caseI), 
the averaged DSW and the VSW (see Fig.~\ref{fig:C2DN}) 
both tend to a single DSW (VSW) once the 
front of the DSW (VSW) passes the front of the RW.

\begin{figure}
\includegraphics[width=3.2in]{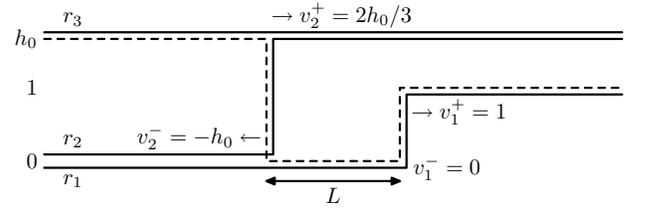} 
\caption{\label{fig:c2wr} The initial data regularization 
of Case~II (\caseII) for $h_0 > 1$, $h_1 = 0$ and $h_2 = 1$; 
the dashed line is the initial condition, $u_0(x)$, 
and the solid lines are $r_1$, $r_2$, and $r_3$.  
The figure also gives the speed of the front and back of the 
DSW and RW at $t=0$.  }
\end{figure}

We can use the one-phase Whitham equations to characterize 
the interaction of the DSW and RW in Case~II (\caseII).  
In this context, 
Whitham theory consists of looking for a fully nonlinear 
single- or multi-phase solution whose parameters 
(amplitude, wave number and frequency) are slowing varying 
with respect to the phase(s) and then deriving new equations 
for the evolution of the slowly varying wave properties.  
The one-phase Whitham equations for (\ref{eq:KdV}) are 
\begin{subequations} \label{eq:1W}
\begin{equation}
  \frac{\partial r_i}{\partial t} 
  + v_i(r_1,r_2,r_3) \frac{\partial r_i}{\partial x} = 0, 
  \qquad i = 1,2,3, 
\end{equation}
where 
\begin{equation} \label{eq:vi}
\begin{aligned}
    v_1 & = V 
      - \frac{2}{3}(r_2 - r_1) \frac{K(m)}{K(m) - E(m)}, \\
    v_2 & = V 
      - \frac{2}{3}(r_2 - r_1) \frac{(1-m) K(m)}{E(m) 
      - (1-m)K(m)}, \\
    v_3 & = V 
      + \frac{2}{3}(r_3 - r_1) \frac{(1-m)K(m)}{E(m)}, 
  \end{aligned} 
\end{equation}
\end{subequations}
$V = (r_1 + r_2 + r_3)/3$, $m = (r_2 - r_1)/(r_3 - r_1),$ 
$K(m)$ is the complete elliptic integral of the first kind, 
and $E(m)$ is the complete elliptic integral of the second 
kind~\cite{Gurevich1974}.  
Then, the asymptotic solution is 
\[ u_a(x,t) \approx r_1 + r_2 - r_3 + 2(r_3 - r_1) 
  \dn^2 \left( \theta; m \right), \]
where $\theta_x = \kappa$, $\theta_t = -\omega = -\kappa V$, 
$\kappa = \sqrt{(r_3 - r_1)/(6\epsilon^2)}$, 
and $r_i$ are slowly varying functions of $x$ and $t$.
We can make a global dispersive regularization for the 
initial value problem (\ref{eq:KdV}) and (\ref{eq:uIC}) 
by choosing appropriate initial data for the $r_i$ 
\cite{Hoefer2006,regTheory} which result in a global solution.
A global dispersive regularization of Case~II (\caseII) 
is shown in Fig.~\ref{fig:c2wr}; 
the $r_i$ are taken to be nondecreasing,  
$r_i(x,0) < r_{i+1}(x,0)$ and $\bar{u}_a(x,0) = u(x,0)$ 
for all $x \in \mathbb{R}$. 

\begin{figure}
\halign{\hfil # \hfil & \hfil # \hfil\cr
  (a) \quad $t = 50$ & (b) \quad $t = 50$ \cr
  \includegraphics[width=1.65in]{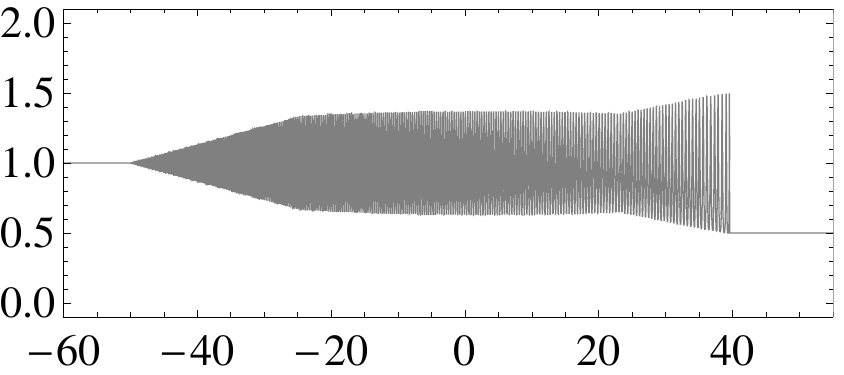} & 
  \includegraphics[width=1.65in]{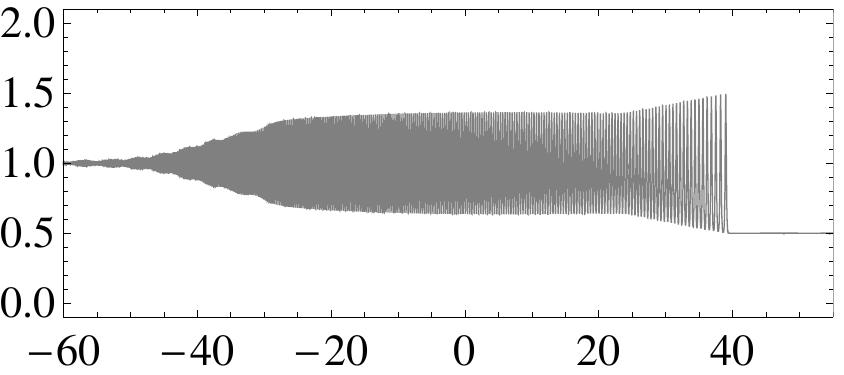} \cr}
\caption{\label{fig:c2dnw} Plot 
(a) shows the Whitham approximation and (b) direct numerics 
of the solution of (\ref{eq:KdV}) for Case~II (\caseII) with  
the same initial condition as Fig.\ \ref{fig:C2DN}.}
\end{figure}

In order to study the interaction we evolve 
the $r_i$ numerically.  
A simple and effective method for evolving the $r_i$ is to 
discretize the initial data regularization along the dependent 
variable, $r_i$, and then compute the shift in $x$ of each data 
point using (\ref{eq:1W}).  
Fig.~\ref{fig:c2dnw} compares a numerically evolved 
Whitham approximation with direct numerics for 
Case~II (\caseII); 
the first order Whitham approximation does not capture the small 
quasi-periodic modulations in the tail because they are 
higher order effects.  
Both direct numerics and the Whitham approximation agree and 
show that for large enough time, 
the amplitude of the tail in Cases II (\caseII) 
is proportional to $t^{-1/2}$;   
this is typical of a uniform linear wave train when 
the total energy remains constant 
(cf.\ \cite{Whitham1965}) and was observed in the context 
of a depression initial condition in \cite{El2002}.  


\medskip

Multiphase Whitham theory is more complicated 
than one-phase Whitham theory and dates back 
to 1970~\cite{Ablowitz1970}; 
multiphase Whitham equations were developed 
for the KdV equation in \cite{Flaschka1980}. 
The interaction of two DSWs from certain step-like data 
was recently analyzed in \cite{Hoefer2007} 
for the nonlinear Schr\"odinger equation.  
The one- and two-phase regions and 
the averaged solution in Case I (\caseI) are found by 
numerically evolving the two-phase Whitham equations 
for the KdV (see \cite{Levermore1988}), 
\begin{equation}
\frac{\partial r_i}{\partial t} + v_i(r_1,\dotsc,r_5) 
  \frac{\partial r_i}{\partial x} = 0, \qquad 
  i = 1,2,\dotsc,5, 
\end{equation}
where 
$v_i = (2r_i^3 - \chi r_i^2 - \beta_1r_i - \beta_2)/(
r_i^2 - \alpha_1r_i - \alpha_2),$ 
$\chi = \sum_{j=1}^{5} r_j$, and 
$\alpha_1$, $\alpha_2$, $\beta_1$ and $\beta_2$ 
are solutions of 
\[  \begin{bmatrix} 
      I_1^1 & I_1^0 \\ I_2^1 & I_2^0 
    \end{bmatrix}
    \begin{bmatrix} \alpha_1 \\ \alpha_2 \end{bmatrix} = 
    \begin{bmatrix} I_1^2 \\ I_2^2 \end{bmatrix}, \quad
    \begin{bmatrix} 
      I_1^1 & I_1^0 \\ I_2^1 & I_2^0 
    \end{bmatrix}
    \begin{bmatrix} \beta_1 \\ \beta_2 \end{bmatrix} = 
    \begin{bmatrix} 
      2I_1^3 - \chi I_1^2  \\ 2 I_2^3 - \chi I_2^2 
    \end{bmatrix}, \]
with  
\begin{equation} \label{eq:hyperelliptic}
  I_j^k = \int_{r_{2j-1}}^{r_{2j}} \frac{\xi^k}%
  {\sqrt{\prod_{i=1}^5 (\xi - r_i)}}\,d\xi. 
\end{equation}


\medskip

\begin{figure}
\halign{\hfil # \hfil & \hfil # \hfil\cr
  (a) \quad $t = 2$ & (c) \quad $t = 2$ \cr
  \includegraphics[width=1.65in]{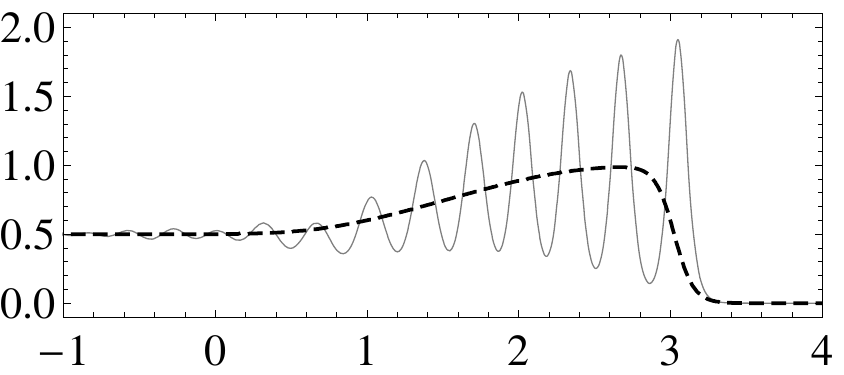} & 
  \includegraphics[width=1.65in]{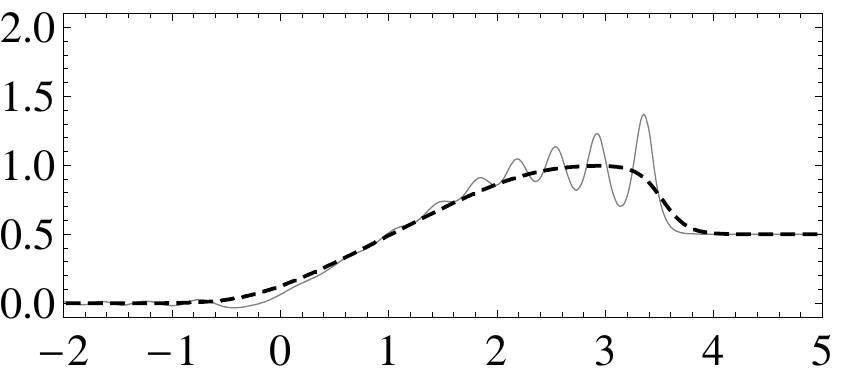} \cr
  (b) \quad $t = 50$ & (d) \quad $t = 50$ \cr
  \includegraphics[width=1.65in]{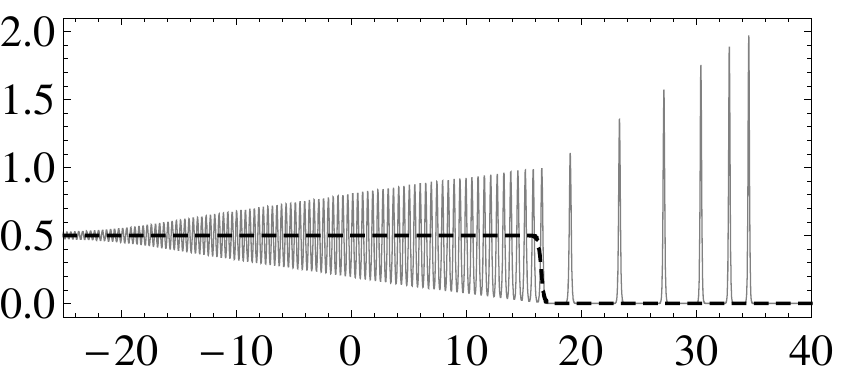} & 
  \includegraphics[width=1.65in]{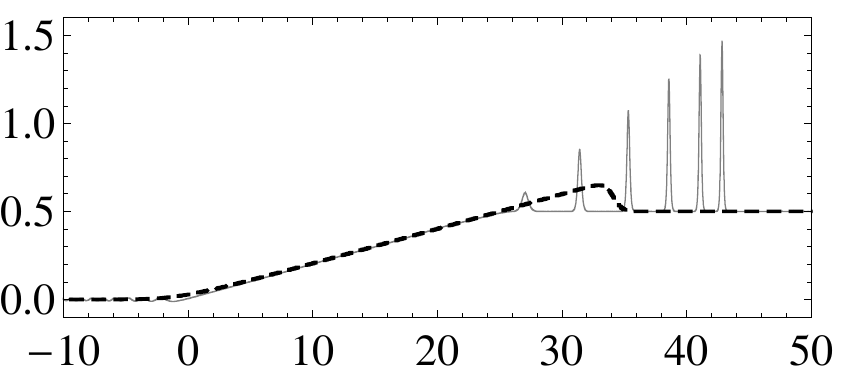} \cr}
\caption{\label{fig:c45} Plots of Cases (a) \& (b) III (\caseIII) 
with $h_0 = 0.5$, $h_1 = 1$, $h_2 = 0$ and (c) \& (d) V (\caseV) 
with $h_0 = 0$, $h_1 = 1$, $h_2 = 0.5$, 
where $\epsilon^2 = 0.001$ and $L = 2$.  
There are six solitons in both cases, see (\ref{eq:akzeroes}). }
\end{figure}

In Case~III (\caseIII), a small RW forms on the left and a large 
DSW forms on the right.   
The front of the RW then interacts with the tail of the DSW 
and reduces the amplitude of the waves---%
essentially cutting off the top of the box.  
Since the front speed of the RW is less than 
the front speed of the initial DSW, 
a finite number of solitons can escape the interaction
(see Fig.~\ref{fig:c45}).  
These solitons have no analogue in the VSW solution of Case~III 
(\caseIII).  
We can compute the number of solitons which escape 
using IST theory.  

From IST theory, 
the number of solitons correspond to the time-independent 
number of zeroes of $a(k)$ (which is the number of poles of the 
reflection coefficient $R \equiv b(k)/a(k)$) in 
the upper half $k$-plane. 
Associated with (\ref{eq:KdV}), the data $a(k)$ is defined by 
\begin{align*}
  \phi(x;k) & \equiv a(k) \bar{\psi}(x;k) + b(k) \psi(x;k), \\
  \bar{\phi}(x;k) & \equiv \bar{a}(k) \psi(x;k) 
    + \bar{b}(k) \bar{\psi}(x;k),
\end{align*}
corresponding to the eigenfunctions,
\begin{gather*}
  \phi(x;k) \sim e^{-i k_0 x}, \qquad 
    \bar{\phi}(x;k) \sim e^{i k_0 x}, \qquad 
    \text{as } x \to -\infty, \\
  \psi(x;k) \sim e^{i k_2 x}, \qquad 
    \bar{\psi}(x;k) \sim e^{-i k_2 x}, \qquad 
    \text{as } x \to +\infty, 
\end{gather*}  
which satisfy the Schr\"odinger scattering problem, 
\begin{equation} \label{eq:scatter}
  w_{xx} + w\{ u/6 + k^2 \}/\epsilon^2  = 0.
\end{equation}
The solution of (\ref{eq:scatter}), at $t=0$, is  
\[ w(x) = 
  \begin{cases} 
    A e^{ik_0x} + B e^{-ik_0x}, & x < 0, \\
    C e^{ik_1x} + D e^{-ik_1x}, &  0 < x < L, \\
    E e^{ik_2x} + F e^{-ik_2x}, & x > L,
  \end{cases} \]
where $k_0 = \sqrt{h_0/6 + k^2}/\epsilon$, 
$k_1 = \sqrt{h_1/6 + k^2}/\epsilon$, 
and $k_2 = \sqrt{h_2/6 + k^2}/\epsilon$.  
The eigenfunctions, 
$\phi$, $\bar{\phi}$, $\psi$ and $\bar{\psi}$ are 
determined by requiring that $w$ and $w'$ are 
continuous across $x = 0$ and $x = L$.  
Indeed, 
$\phi$ is found by taking $A = 0$ and $B = 1$ 
and then solving for $C$, $D$, $E \equiv b(k)$, $F \equiv a(k)$, 
so that 
\[  a(k) = e^{ik_2L}\frac{k_0 + k_2}{2k_2} \bigg\{ \cos(k_1 L) 
  - i\frac{k_1^2 + k_0 k_2}{k_1(k_0 + k_2)} \sin(k_1 L) \bigg\}. \]
Since $e^{ik_2L}(k_0 + k_2)/(2k_2) \neq 0,$ 
the zeroes of $a(k)$ occur when 
$\tan(k_1 L) = ik_1(k_0 + k_2)/(k_1^2 + k_0 k_2)$.

It can be shown that the zeroes of $a(k)$ are purely imaginary; 
thus, we let $k = i\kappa$ 
(where $\kappa \in \mathbb{R}$ and $\kappa > 0$).  
For Case~III (\caseIII), where $h_1 = 1 > h_0 = h_*$ and $h_2 = 0$, 
the zeroes of $a(i\kappa)$ occur when 
\begin{equation} \label{eq:akzeroes}
\tan\left( \sqrt{1/6 - \kappa^2}L/\epsilon\right) = 
  \frac{\sqrt{1/6 - \kappa^2}\left(\sqrt{\kappa^2 
    - h_*/6} + \kappa\right)}%
  {1/6 - \kappa^2 - \kappa\sqrt{\kappa^2-h_*/6}}. 
\end{equation}
The number of periods for 
$\sqrt{h_*/6} \le \kappa \le \sqrt{1/6}$
of the RHS of (\ref{eq:akzeroes}),
$L\sqrt{1 - h_*}/(\epsilon \pi \sqrt{6})$, 
is an estimate of the number of solitons.
The number of zeroes determined using (\ref{eq:akzeroes}) 
exactly corresponds to the number of solitons observed 
using direct numerics (for various values of $h_*$, 
$L$ and $\epsilon$)!  

\medskip

\begin{figure}
\halign{\hfil # \hfil & \hfil # \hfil\cr
  (a) \quad $t = 5$ & (b) \quad $t = 400$ \cr
  \includegraphics[width=1.65in]{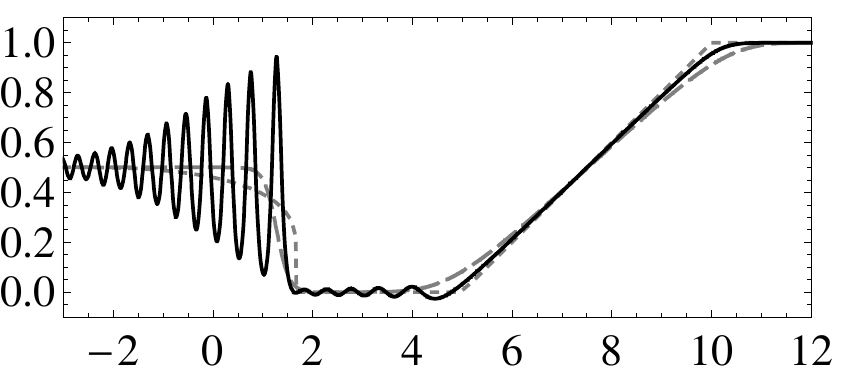} & 
  \includegraphics[width=1.65in]{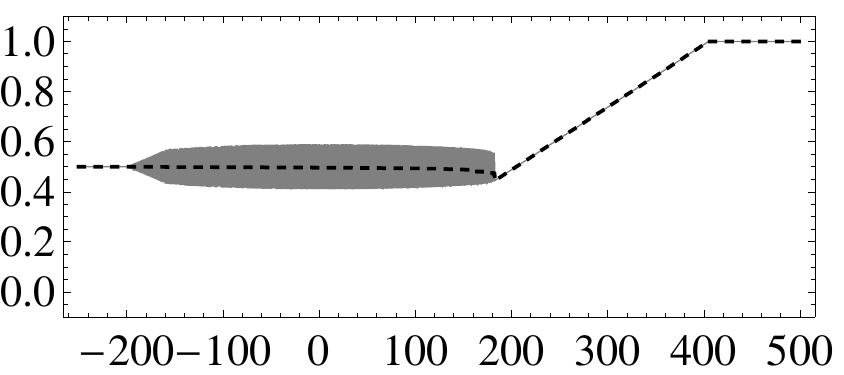} \cr}
\caption{\label{fig:C3DN} 
Plots of the solution of (\ref{eq:KdV}) for Case~IV (\caseIV)
where $\epsilon^2 = 0.001$, $h_0 = 0.5$, $h_1 = 0$, $h_2 = 1$
and $L = 5$.}
\end{figure}

In Case~IV (\caseIV), 
a small DSW forms on the left and a large RW forms on the right 
(see Fig.~\ref{fig:C3DN}a).  
As in Case~II (\caseII), 
the front of the DSW interacts with the trailing edge of the RW 
and decreases the DSW's amplitude and speed.  
Unlike Case~II (\caseII), 
the front of the DSW does not overtake the front of the RW.  
The DSW becomes a small amplitude tail on the left of the RW 
and decreases in amplitude proportional to $t^{-1/2}$ 
(see Fig.~\ref{fig:C3DN}b).  

\medskip

As in Case~III (\caseIII), 
Case~V (\caseV) cannot be completely characterized using 
Whitham averaging.  
For Case~V (\caseV), a large RW forms on the left and a small DSW 
forms on the right; the front of the RW interacts with the tail 
of the DSW and results in a RW and a finite number of solitons. 
The number of solitons corresponds to the number of zeroes of 
(\ref{eq:akzeroes}) where where $h_0 = 0$ and $h_1 = 1 > h_2 = h_*$. 

\medskip

In Case~VI (\caseVI), two rarefaction waves form; 
the small amplitude oscillatory tail 
(see for instance the RW in \ref{fig:C3DN}a) 
of the right RW interacts with the front of left RW; 
the tail of the right and left RW then interact to form a small 
amplitude, modulated, quasi-periodic tail; 
this modulation decreases with time and 
Case~VI (\caseVI) tends to a pure RW for large time.  

\medskip

We numerically solve (\ref{eq:KdV}) and (\ref{eq:Burgers}) using 
an adaptation of the modified exponential time-differencing 
fourth-order Runge-Kutta (ETDRK4) method (see 
\cite{ETDRK4}).  
We use this (sophisticated) numerical method because 
(\ref{eq:KdV}) is very stiff and standard numerical methods
require the time step to be $O(\epsilon^3)$, 
while for ETDRK4 the time step need only be $O(\epsilon)$.  
When this numerical scheme was used to compute a known 
exact solution, it was accurate to more than six decimal digits.

For spectral accuracy when using the ETDRK4 method, 
the initial data must be both smooth and periodic.  
Therefore, we differentiate (\ref{eq:KdV}) with respect to $x$ 
and define $v \equiv u_x$ to get  
$ v_t + (uv)_x + \epsilon^2 v_{xxx} = 0. $
Transforming to Fourier space gives 
$ \hat{v}_t = i\epsilon^2 k^3 \hat{v} - i k \widehat{uv} 
  \equiv \L\hat{v} + \N(\hat{v},t), $
where we define $(\L\hat{v})(k) \equiv i\epsilon^2 k^3 \hat{v}$ 
and $\N(\hat{v},t) = \N(\hat{v}) \equiv -i k \F\{[h_0 
+ \int_{-\infty}^x \F^{-1}(\hat{v}) dx']\F^{-1}(\hat{v})\}$.  
It is important that the integral in $\N$ is computed using 
a spectrally accurate method.   
Moreover, we approximate the initial step data 
with the analytic function 
$ 2w v(x,0) =  (h_2 - h_1)\sech^2[(x-L)/w]
  + (h_1 - h_0)\sech^2(x/w)$, 
where $w$ is small.  
See \cite{ETDRK4} for details about how this 
$\L$ and $\N$ are used to numerically compute the solution 
of (\ref{eq:KdV}).


\medskip

For large time Case~I (\caseI) and~II (\caseII) go to a single 
DSW, while Case~IV (\caseIV) and~VI (\caseVI) go to a single RW; 
this is consistent with VSW theory.
However, unlike VSW theory, Case~III (\caseIII) and~V (\caseV) 
form a finite number of solitons in addition to the DSW or~RW, 
respectively.  
Moreover, unlike VSW theory, Case~I (\caseI) exhibits a 
transient two-phase region and Case~II (\caseII) and~IV 
(\caseIV) have a small amplitude tail which decays at a rate 
proportional to~$t^{-1/2}$.  

\begin{acknowledgments}
This work was partially supported by NSF DMS-0604151, 
DMS-0803074, Air Force Office of Scientific Research FA-9550-06-1-0237, 
and NDSEG fellowship.  
\end{acknowledgments}

\frenchspacing

\end{document}